\documentclass[12pt]{article}
\usepackage{graphicx}
\newcommand{\ber}{\begin{eqnarray}}
\newcommand{\eer}{\end{eqnarray}}
\newcommand{\bea}{\begin{equation}}
\newcommand{\eea}{\end{equation}}
\newcommand{\del}{\partial}
\begin{document}
%\maketitle
\title{\bf On the path integral simulation of space-time fractional Schr\"{o}dinger equation with time independent potentials \\}
\author{\bf Sumita Datta $^{1,2}$\\
$^1$ Allinace School of Applied Mathematics, Alliance University,\\ Bengaluru 562 106, India\\
$^2$ Department of Physics, University of Texas at Arlington,\\Texas 76019, USA\\
{\bf Radhika Prosad Datta}\\
Indian Institute Of Foreign Trade(IIFT), Kolkata Campus\\
Kolkata-700017,India\\}
 
\maketitle
\begin{abstract}
In this work a Feynman-Kac path integral method based on L$\acute{e}$vy measure has been proposed for solving the Cauchy problems associated 
with the space-time fractional Schr\"{o}dinger equations arising in interacting systems in fractional quantum mechanics. The Continuous Time Random Walk(CTRW) model is used to simulate the underlying  L$\acute{e}$vy process- a generalized Wiener process. 
Since we are interested  to capture the lowest energy state of the quantum systems, we use Pareto distributions as opposed to Mittag-Leffler random variables,  which are  more suitable for finite time. Adopting 
the CTRW model we have been able to simulate the space-time fractional diffusion process with comparable simplicity and covergence rate as in the case of standard diffusion processes. We hope this paves an elegant way to solve space-time diffusion equations numerically through Fractional Feynman-Kac path integral technique as an alternative to fractional calculus. 
\end{abstract}
\newpage
\section{Introduction}
Fractional calculus enables one to generalize standard diffusion equation by replacing the space and time indices with non-integers. The fundamental solutions of these generalized diffusion equations are intimately connected to some non-Gaussian probability distributions[1] which occur in many fields ranging from finance to physics. This has led to the growing interest in fractional calculus to enlarge the domain of stochastic models by non-Gaussian distributions. For solving the frctional eigenvalue problems in fractional quantum mechanics the most typical approach would be to solve fractional Schr\"{o}dinger equation by classical Feynman path integrals[2-4]. 
Solving eigenvalue problems related to fractional Schr\"{o}dinger equations with local potential is usually difficult because of the intrinsic nonlocality of the Riesz derivatives[5] in the Hamiltonian. Because of the two fold complexities of fractional derivatives and quantum mechanics not much progress has been been made in the numerical simulation work even on the  fractional quantum problems with simple time independent potentials.\\

Eventhough the literature shows the following works on similar lines there is a  lot of limitations and controversies regarding the approaches in fractional quantum mechanics.
The connection between the fractality of L$\acute{e}$vy paths and space-fractional quantum mechanics was established in the pioneering
 work of Laskin[6] and later on, it was extended by him for space-time fractional quantum mechanics in [9]. Around the year 2000, Laskin came up with a series of seminal papers[6-9] which contain primarily the eigensolutions of space-fractional Schr\"{o}dinger equation for  different quantum systems by L$\acute{e}$vy path integral approach. Subsequently the concept of time fractional Schr\"{o}dinger equation was initiated and implemented by Naber[10]. Dong[11] provided the solution of space-time fractional equation for the infinite square well potential by following Laskin's L$\acute{e}$vy path integral approach. Iomin[12,13] solved the infinite square well problem by using the topological constraints. Wang  and Xu[14] came up  with an exact solution to the space-time fractional Schr\"{o}dinger equation with infinite square well potential whereas Dong and Xu[15] solved the space-time fractional Schr\"{o}dinger equation with delta function potential. The Cauchy problem for a space-time fractional Schr\"{o}dinger equation with delta function potential was solved by Saberhaghparvar et al[16] whereas the space fractional Schr\"{o}dinger equation with delta function potential was  solved by Oliveria et al[17]. Time fractional linear and nonlinear Schr\"{o}dinger equation were also looked at in ref[18-19] and [20] respectively.\\ 

In the above list, some  work was done within semiclassical approximation and others with  local  approaches  which are not well suited for fractional quantum mechanics.
For the above reasons and to fill the gap for the numerical  work we propose the numerical simulation of the diffrent fractional quantum mechanical systems by Feynman-Kac(FK) path integral approach[2,3,4] and  generalized Feynmna-Kac(GFK) path integral approach[21-26] with L$\acute{e}$vy measure[27]. This we call our version of `L$\acute{e}$vy path integral method' which 
 was inspired by our earlier work[24-26] on standard diffusion problem with FK and GFK path integral formalism  and the extension of FK path integral formalism for space fractional Schr\"{o}dinger Equation developed in ref[27,28].\\                                                                       
            
In this paper we solve the eigenvalue problem associated with  space-time fractional  Schr\"{o}dinger Equations in fractional quantum mechanics using a Quantum Monte Carlo method with specific appliaction to the problem of ref[29,30,31] where  
it was shown that a Bose-Einstein condensate with a  strong correlation amongst the constituents satisfies a generalized Gross-Pitaevskii[32] equation  with momentum and energy operator with fractional powers different from space index $\alpha=2$ and time index $\beta=1$ respectively involved in standard diffusions. 
The  equation mentioned above with fractional space and time indices can be solved exactly using Fox's H function[33,34].
We want to have a stochastic solution to it. Another L$\acute{e}$vy distribution has been identified in a strongly correlated system in ref[35].\\

 Here the simualtion of the stochastic processes is done with Continuous Time Random Walk(CTRW) model[36]. CTRW is a random walk in which both step lengths and the waiting times are distributed according to a L$\acute{e}$vy disbtribution[37].  While a lot of analytical work[38-48] has been done in connecting the integral equation for CTRW to space time fractional diffusion equation, not much attempts have been to do the eigenvalue calculations in fractional quantum mechanics. In our work the nonlocal fractional Laplacian gets absorbed in so called L$\acute{e}$vy measure. We just sample the time independent potential as a function of the random process with L$\acute{e}$vy measure which makes it a local approach. 
 We have done a few bench mark eigenvalue calculations for  harmonic oscillator and system of paricles interacting with strong delta function potential in the space fractional case using FK and GFK path integral method.  As for the space-time fractional case we do not have any bench marks to compare, with we have identified some of the traits in the density profiles of the interacting systems under consideration as the hallmarks of power law distribution; a fat tail in the density profile and depletion in the density in the space-time fractional case,  to name a few. We establish that this way we make an interesting contribution to the field of fractional quantum mechanics.\\

This work is set out as follows. In Section 1 we introduce the CTRW random walk approach to the fractional diffusion equation, which allows us to establish a relation between fractional diffusion equation and power law distribution. In Section 2 we describe some of the key concepts required to understand  different path integration techniques derived in Section 3. Further in Section 4 we discuss the numerical schemes involved in the work for solving  the eigenvalue problems and fractal aspect related to the space-time fractional diffusion equations. We discuss the analytical and numerical results in Section 5. Finally in Section 6 we discuss the implications of our findings and propose some directions for future work.   
\section{Notions and notations}
{\bf I. Standard diffusion equation vs space-time fractional diffusion equation.}
Standard diffusion equation with initial condition can be written  in $Schr\ddot{o}dinger$ representation as follows:
\ber
\frac{\partial \psi(t,\vec{x})}{\partial t}=D\frac{{\partial}^2\psi(t,\vec{x})}{\partial x^2} \\ \nonumber
\psi(0^+,x)=\delta(x)
\eer
Using Montroll-Weiss equation[36] the above classical heat equation can be generalized to a space-time fractional equation:
\bea
\frac{{\partial}^{\beta}\psi(t,\vec{x})}{{\partial t}^{\beta}}=D_{\alpha}\frac{{\partial}^{\alpha}\psi(t,\vec{x})}
{{\partial |x|}^{\alpha}}
\eea
where $0< \alpha \le 2$ and $ 0 < \beta \le 1$.\\
$\frac{{\partial}^\beta}{\partial t^\beta}$ and $ \frac{{\partial}^\beta}{{\partial |x|}^{\alpha}} $ are called the Caputo derivative[49] and Riesz derivative[5] respectively.\\
{\bf II. Extension of fractional heat equation in the interacting system by Feynman-Kac formula:} 
{\bf Proposition:} Let the $Schr\ddot{o}dinger$ semigroup[50] $T(t)=e^{tH_0}$ be defined by $T(t)f(x)=E_xf(X(t))$ for f in $L$ where L is the Banach space of bounded real valued functions with generator of the semigroup $ A=H_0=\frac{{\Delta}^\alpha}{2}$. Then 
$ \tilde T(t)f(x)=E_x {e^{-\int_{0}^{t}V(X(s))ds}f(X(t)} $ defines a semigroup with inifinitesimal generator $\tilde A$ such that 
$\tilde Af=Af(x)+V(x)f(x) $ and $D(\tilde A)=D(A)$. Hence $ \psi(t,\vec{x})=\tilde T(t)f(x) $ solves $\frac{{\partial}^{\beta}\psi(t,\vec{x})}{{\partial t}^{\beta}}=D_{\alpha}\frac{{\partial}^{\alpha}\psi(t,\vec{x})}
{{\partial |x|}^{\alpha}}+V\psi(t,\vec{x})$.\\ 
{\bf  III. Introduction of Importance sampling for the above generalization with Generalized Feynman-Kac path integration:}
Now let us consider the Stochastic Differential equation(SDE)
$dY(t)=b(X(t))dt+dX(t),X_0=x$. Note that the infinitesimal generator of $Y(t)$ is given by 
$\mathcal{L}\psi(x)=-{\Delta}^{\frac{\alpha}{2}}+b(x)\nabla \psi(x)$
Let $f$ and $V$ be bounded continuous functions on $R^d$. Then the FK semigroup corresponding to the above SDE is given by
$P_t^Vf(x)= E_x [{e^{-\int_{0}^{t}V(X(s))ds}f(X(t)}]$. Then under certain additional assumptions, the function $\psi(t,\vec{x})=P_t^Vf(x),t\ge 0,x\in R^d$ provides a probabilistic representation for the following equation[51]:
\ber
\frac{{\partial}^{\beta}\psi(t,\vec{x})}{{\partial t}^{\beta}}  
=\mathcal{L}\psi(t,\vec{x})-V(x)\psi(t,\vec{x}) \\ \nonumber 
=-{\Delta}^{\frac{\alpha}{2}}+b(x)\nabla \psi(x)-V(x)\psi(t,\vec{x}) \\ 
\psi(0,x)=f(x) \nonumber
\eer
\section{Standard Path integral formalism }
We need to take a resort to Quantum Monte Carlo method based on path integration(FK) to solve the eigenvalue problem of the quantum systems.
Metropolis and Ulam[52] were the first to exploit relationship between the Schr\"{o}dinger equation for imaginary time and the random-walk solution of diffusion equation.
\subsection{Feynman-Kac path Integral representation}
Let us consider the initial value problem
\ber
& & i\frac{\del{\psi}(x,t)}{\del{t}}=-(\frac{\Delta}{2}+V){\psi}(x,t)\nonumber \\
& & {\psi}(x,0)=g(x)
\eer
where $x\in R^d$.
Applying Trotter[50,53] product the solution of the above equation can be written as 
\bea
\psi(x,t)=\lim_{n\rightarrow \infty}\int e^{iS_n(x_1,x_2,.......,t)}f(x_n)dx_n.............dx_1
\eea
The integration is done on $R^n$  and $ S_n(x_1,x_2,.......,t)=\sum_{j=1}^n\frac{t}{n}[\frac{1}{2}\frac{|x_j-x_{j-1}|^2}{t/n}-V(x_j)]$
Now Eq(5) can be written  as
\ber
\psi(x,t)=\lim_{n\rightarrow \infty}\int e^{iS_n(x_1,x_2,.......,t)}f(x_n)d{\mu}_n
=\int e^{iS(x_1,x_2,.......,t)}f(x)d\mu
\eer
provided $d{\mu}_n=(\frac{2{\pi}it}{n})^{{-3n}/2}dx_n.............dx_1$ converges to $d\mu$ and $S_n$ converges to S.
Note that now the integration in $R^{\infty}$. Unfortunately, one cannot justify the existence of S and $\mu$ because of the following reasons:\\
(a) most paths are not differetiable as a function of time.\\
(b) $d{\mu}_n=(\frac{2{\pi}it}{n})^{{-3n}/2}dx_n.............dx_1$ does not exist as $n\rightarrow \infty$.\\
To overcome the difficulty associated with the non-existence of complex valued $\mu(x)$ one replaces t by -it and obtains
\ber
& & \frac{\del{\psi}(x,t)}{\del{t}}=(\frac{\Delta}{2}-V){\psi}(x,t)\nonumber \\
& & {\psi}(x,0)=g(x)
\eer
where $x\in R^d$.
The solution of the above equation in Feynman represetation[54,55,56] can be given by 
\bea
\psi(x,t)=(e^{itH}f)(x)=\int\prod_{j=1}^n e^{{-t/n}V(x_j)}f{\tilde{\mu}}_{x_0}^n
\eea
where unlike the previously considered complex case
\bea
d{\tilde{\mu}}_{x_0}^n=dx_1.............dx_n k(x_0,x_1;t/n)........k(x_{n-1},x_n;t/n) 
\eea
with $k(x,y,t)=(2{\pi}t)^{-n/2}e^{-{(x-y)^2}/{2t}}$, is a real valued measure which has a limit as $n\rightarrow \infty$. Namely ${\tilde{\mu}}_{x_0}^n
$ can be identified with a probability measure, i.e., Wiener measure[57]. In terms of the propagator the solution of eq(8) can be written as[2] $\psi(x,t)=\int k(x,y,t)\psi(y)dy $ where $ k(x,y,t)$ is the time dependent propagator. 
The  solution provided in equation(8) can be written in Feynman-Kac
representation as
\bea
{\psi}(x,t)=E_{x}[e^{-\int_{0}^{t}V(X(s))ds}g(X(t))]
\eea
where $E_x$ is the expected value of the random variable $g(X(t))$ provided $X(0)=x$
for $V\in K_{\nu}$, the Kato class of potential[58].
A direct benefit of having the above representation is to recover the lowest energy eigenvalue of the Hamiltonian
$H=-\frac{\Delta}{2}+V $ for a given symmetry by applying the large deviation principle of Donsker and Varadhan as follows[59]:
\bea
{\lambda}_1=-\lim_{t\rightarrow \infty}\frac{1}{t}lnE_x[{e^{-\int_{0}^{t}V(X(s))ds}g(X(t))}]
\eea 
Density can be calculated as 
\bea
\rho=|\psi(x,t)|^2
\eea
\subsection{Generalized Feynman-Kac path integration}
The above
representation(Eq[10]) suffers from poor convergence rate as the underlying diffusion process-Brownian motion(Wiener Process) is non-recurrent. So it is necessary to use a representation which employs a diffusion which unlike Brownian motion[21,23], has a stationary distributions.
For any twice differentiable $\phi(x)>0$ define a new potential U which is a
perturbation of the potential V as follows[21,23].
%\bea
%U(x)=V(x)-\frac{1}{2}\frac{\del \phi}{\del x}
%\eea
\bea
U(x)=V(x)-\frac{1}{2} \frac{\Delta \phi(x)}{\phi(x)}\,.
\eea
Also consider
\bea
w(x,t)=E_{x}[e^{-\int_{0}^{t}U(Y(s))ds}h(Y(t))]
\eea
which is Feynman-Kac solution to
\ber
\frac{\del w(x,t)}{\del t}
& & =\frac{1}{2}\Delta w(x,t)+\frac{\nabla \phi(x)}{\phi (x)}\nabla w(x,t)-U(x) w(x,t)\\
 & & =-Lw(x,t)
\nonumber\\
& & w(x,0)=h(x)\,,
\nonumber\\
\eer
where $h$ is the initial value of $w(x,t)$.
The new diffusion $ Y(t) $ has an infinitesimal generator $A=\frac{\Delta}{2}+\frac{\nabla \phi}{\phi}\nabla $, whose adjoint is $A^{\star}(\cdot)=\frac{\Delta}{2}-\nabla(\frac{\nabla \phi}{\phi}(\cdot))$. Here ${\phi}^2(x)$ is a stationary density of $ Y(t) $, or equivalently, $A^{\star}({\phi}^2)=0$.
To see the connection between $ w(x,t) $ and $ \psi(x,t) $, observe that for $g=1 $ and $h=1$,
\bea
w(x,t)=\frac{{\psi}(x,t)}{\phi(x)}\,,
\label{former_23}
\eea
because $ w(x,t) $ satisfies Equation (15). The diffusion $ Y(t) $ solves the following stochastic differential equation:
$dY(t)=\frac{\nabla \phi(Y(t))}{\phi(Y(t))}+dX(t)$\,.
$V(Y(s))$ is summed over all the time steps and $e^{-V(Y(s))}$ is summed over all the trajectories.
The presence of both drift and diffusion terms in the above expression enables the trajectory to be highly localized. As a result, the important regions of the potential are frequently sampled and Eq(10) converges rapidly.
The expectation value for the other properties can be evaluated as follows[21,25]:\\
\bea
{\langle Y|A|Y\rangle}=
\frac{\lim_{t\to\infty}\int dY(t)A(Y(t))e^{-\int[{V}_p(Y(s)]ds}}
{\int dY(t)e^{-\int[{V}_p(Y(s)]ds}}\,.
\label{former_13}
\eea
Moreover, the eigenvalue problem for the stationary $Schr\ddot{o}dinger$ equation 
\bea
-H\psi=\frac{1}{2}\Delta\psi-V\psi=-\lambda\psi
\eea 
and
\bea
-L\tilde{\psi}=-\frac{1}{2}\Delta{\tilde{\psi}}+\frac{\nabla \phi}{\phi}\nabla{\tilde{\psi}-U\tilde{\psi}}
=-\lambda\tilde{\psi}
\eea
are related as follows:
\bea
\tilde{\lambda}=\lambda
\eea
\bea
\tilde{\psi}=\frac{\psi(x)}{\phi(x)}
\eea 
In GFK we generate a diffusion process with the aid of a twice differentiable non-negative function  ${\phi}_0(x)$ by considering the following Hamiltonian:
\bea
H_0=-\frac{\Delta}{2}+U_0
\eea
\bea
U_0=e_0+\frac{1}{2} \frac{\Delta {\phi}_0(x)}{{\phi}_0(x)}
\eea
Here ${\phi}_0$ has been chosen as a trial function associated with the symmetry of the problem.
The main reason for introducing $H_0$ with a localized diffusion process is that it possesses a stationary distribution, unlike the nonlocalized Brownian motion process that escapes to infinity[60]. This way one achieves the so called important sampling goal in the actual numerical computation, which thereby reduces the computing time considerably.
Now let us decompose the Hamiltonian of the quantum mechanical problem into two parts: $H=H_0+V_p$
In terms of the energy associated with the trial function,$e_0$, we now define a new perturbed potential
\bea
V_p(x)=V-U_0
=V-e_0-\frac{1}{2} \frac{\Delta {\phi}_0(x)}{{\phi}_0(x)}
\eea
\subsection{Continuous Time Random Walk(CTRW)}
In this set up a Continuous Time Random Walk (CTRW)[36,38-48] trajectory can be defined as
\bea
X(t)=X_0+\sum_{n=1}^{N(t)}\Delta{X_n}
\eea
where $N(t)=sup\{n|t_n\le{t}\}$. The the CTRW experiences a jump $\Delta{X_n}=\Delta{X_n}(t_n)=X(t_n)-X(t_{n-1})$ at the end of nth sojourn or waiting time,$\Delta{t_n}=t_n-t_{n-1}$. $\Delta{X_n}$ and $\Delta{t_n}$ are iids. The probability density for the process to take value X at time t
is given by
\bea
p(x,t)=\sum_{n=0}^{\infty}p(n,t)p_n(x)
\eea
where $p_n(X)$ is the probability for the process of taking the value X after n jumps and $ p(n,t)$ is the probability of having n jumps after time t.  Here we assume that  $p_n(X)$ and $ p(n,t)$ are independent and $ \eta(\Delta t)$ and $f(\Delta X)$ are the distributions of $\Delta t$ and $\Delta X$ respectively. The Laplace transform of $ \eta(\Delta t)$ reads as
$\tilde {\eta}(s)=\int_{0}^{\infty} d\tau e^{-\Delta t s}\eta{\Delta t}$
and the Fourier transform for $f(\Delta x)$ ca be read as
$\hat{f}(k)=\int_{-\infty}^{\infty}d(\Delta x)e^{ik\Delta x}f(\Delta x)$ 
In the Fourier-Laplace  Eq(27) can be expressed as 
\bea
\hat{\tilde{p}}(k,s)=\frac{1-\tilde{\eta}(s)}{s}\frac{1}{\tilde{\eta}(s)\hat{f}(k)}
\eea
The above equation is known as the MONTROLL-WEISS(MW) equation and by taking the Laplace-Fourier transform of MW equation one gets the
the integral equation of the CTRW as follows:
\bea
p(x,t)=\delta(x)\Phi(t)+\int_0^t\{\int_{-\infty}^{\infty}f(x-x^{\prime})p(x^{\prime},t^{\prime})dx^{\prime}\}\phi(t-t^{\prime})dt^{\prime}
\eea
Now under the power law assumption of the jump and waiting time and proper scaling of space and time it can be shown that the $p(x,t)$ actually tends weeakly to the solution of the Cauchy problem 
\ber
\frac{{\partial}^{\beta}\psi(t,\vec{x})}{{\partial t}^{\beta}}=D_{\alpha}\frac{{\partial}^{\alpha}\psi(t,\vec{x})}
{{\partial |x|}^{\alpha}}
\psi(0^+,x)=\delta(x)
\eer
Here where $0< \alpha \le 2$ and $ 0 < \beta \le 1$. This forms the basis of our path integral Monte Carlo simulation of space-time fractional Schr\"{o}dinger equation. 
\subsection{Fractional Feynman-Kac path integral}
Our derivation of this section is based on the the previous work on the fractional generalization of FK path integration[27,28] 
Following the derivation in the one can write the Feynman-Kac functional integral as
\bea
\psi(x,t|y,0)=\int_{C(0,t)} d{\tilde{\mu}}_{x_0}^n e^{-\int_0^t V(x(s)ds}
\eea
where
\ber 
d{\tilde{\mu}}_{x_0}^n=dx_1.............dx_n k(x_0,x_1;t/n)........k(x_{n-1},x_n;t/n)
=\lim_{n\rightarrow \infty}\prod_{j=1}^n k(\Delta {x_i},\Delta {t_j}) dx_n
\eer
Following ref[59] and considering the Fourier transform for $k(x,t)$ one can show that for the space index $\alpha=2$  
\ber
\psi(x,t|y,0)=\int_{R^{2n}} dx_1 dp_1.....dx_n dp_n e^{\sum_{j=1}^n (ip_j\Delta {x_j}-\Delta{t_j}[D_2 p_j^2-V(x_j)])}
\eer 
One can then generalize the above expression for $\alpha \le 2$ as
\ber
\psi(x,t|y,0)=\int_{R^{2n}} dx_1 dp_1.....dx_n dp_n e^{\sum_{j=1}^n (ip_j\Delta {x_j}-\Delta{t_j}[D_{\alpha} p_{j}^{\alpha}-V(x_j)])}
\eer
Now by defining a measure 
\bea
d{\mu}^L=dx_1.............dx_n \int_{R^n} dp_j e^{\sum_{j=1}^n (ip_j\Delta {x_j}-\Delta{t_j}[D_{\alpha} p_{j}^{\alpha}])}
= \lim_{n\rightarrow \infty}\prod_{j=1}^n k(\Delta {x_i},\Delta {t_j}) dx_n
\eea
\bea
\psi(x,t|y,0)=\int_{C(0,t)} d{\mu}^L e^{-\int_0^t V(x(s)ds}
\eea
for $\alpha \le 2$ and it is called a  L$\acute{e}$vy measure.
For space fractional Schr\"{o}dinger equation($\alpha \le 2 $ and $\beta \le 1$ ) the propagator in equation(35) is denoted as 
$k_{\alpha,1}(x,t)$. For space-time fractional equation it  can be written as[1] $ k_{\alpha,\beta}(x/{t^\kappa})$ where $\kappa={\beta/\alpha}$ 
\newpage
\section{Systems to solve by fractional path intgral}
1) Fractional harmonic oscillator[7] \\
Here we solve $H\psi=E\psi$ with
\bea
H_{\alpha,\gamma}=D_{\alpha}(-\Delta)^{\alpha/2}+q^2x^{\gamma}
\eea
Here q is a constant and $\gamma$ is an exponent of x. For harmonic oscillator $\gamma=2 $.
2) Strongly correaled $ \delta $ function problem in ref[29,30]
\bea
%[\hat{E}^{\beta}]
[\hat{E}^{\beta}-D_{\alpha}({\hat{\vec{P}}}^2)^{\frac{\alpha}{2}}-\frac{\delta+1}{4} \\ \nonumber
g_c|\psi(t,\vec{x})|^{\delta-1}]\psi(t,\vec{x})=0
\eea
For $\delta=1$ the above equation with a strong delta function potential can be approximated as
\bea
[\hat{E}^{\beta}-D_{\alpha}({\hat{\vec{P}}}^2)^{\frac{\alpha}{2}} \\ \nonumber
-\frac{g_c \delta(\vec{x})}{2}]\psi(t,\vec{x})=0
\eea
Using the energy momentum operator explicitly, $t=-it$ and $V=\frac{g_c \delta(\vec{x})}{2}$ we get
\bea
\frac{{\partial}^{\beta}\psi(t,\vec{x})}{{\partial t}^{\beta}}=D_{\alpha}\frac{{\partial}^{\alpha}\psi(t,\vec{x})}
{{\partial |x|}^{\alpha}}+V\psi(t,\vec{x})
\eea

\subsection{Numerical implementation of FK path integration in Fractional Quantum Mechanics}
 Now we first make a  connection between the path integral represenation of a  standard diffusion equation(with space index $\alpha=2$ and time index $\beta=1$  to that of a fractional diffusion equation( with non integral space and time indices) and then relate Continuous Time Random walk(CTRW) and fractional  Schr\"{o}dinger Equation. In the classical Feynman-Kac path integral approach the eigenvalues can be computed using Wiener measure(i.e., the probability measure on the space of continuous functions) which uses Brownian Random Walk and provides a rigorous mathematical justification unlike the ordinary Feynman path integration. But the underlying stochastic processes in the case of fractional Schr\"{o}dinger Equation are non-Gaussian as opposed to Brownian motion. In the case of a Brownian motion the 'Central Limit Theorem' holds i.e., the sum of a number of independent and identically distributed random variables with finite moments will tend to a normal variable as ${n\rightarrow \infty}$. In the case of a fractional Schr\"{o}dinger Equation one can still come up with a very efficient yet simple quantum Monte Carlo algorithm due to the 'Generalized Central Limit Theorem'[61] which states that the sum of a number of random variables with a power law tail (Paretian tail) distribution $P\propto \frac{1}{x^{\alpha+1}}$ therby having infinite variances tend to a stable distribution namely 'L$\acute{e}$vy Stable distribution'. This makes the numerical basis of our FK path integral approach to fractional Schr\"{o}dinger Equation. While a lot of analytical work[CTRW] has been done in connecting the integral equation for CTRW to space time fractional diffusion equation, not much attempts have been to do the eigenvalue calculations in fractional quantum mechanics.   

Now to simulate this CTRW by FK path integration method we must note that for this CTRW the probability distributions  of waiting time and jumps have fat tails charecterized by 
 power laws with exponent 0 and 1 for the waiting times, between 0 and 2 for the jumps. To numerically achieve this we  generate the Levy stable distribution using Pareto distribution of second kind[62]:
\bea
f(x)=\frac{\alpha}{(1+x)^{(\alpha+1)}}
\eea
The scaling law used to simulate the  trajectories in the case of  
(i) Normal diffusion $ \Delta x =(\Delta t)^{1/2}$ \\
(ii)Space fractional[70,71] equation $ \Delta x =(\Delta t)^{1/\alpha}$ \\
(iii)Space-time fractional[1] equqtion $ \Delta x =(\Delta t)^{\beta/\alpha}$ \\
The random numbers were generated using the algorithm in [63]. We do not use Mittag-Leffler[64] random number. 
The energy eigenvalues are calculated using Eq(34) and Eq(10) for Feynman-Kac method and Eq(34) and Eq(14)for Generalized Feynman-Kac method respectively. The Energy eigenvalue obtained by Fractional path integration has been compared with those given by the analytical expression provided in the literature.
The analytical expression for energy eigenvalue for the space fractional case with space index  $0< \alpha \le 2$ and $  \beta= 1$ is given as follows:\\
1) For $\delta$ function potential[17]
\bea
E_{\delta}=(\frac{g \csc{\pi/\alpha}}{\alpha \hbar {D_{\alpha}}^{1/\alpha}})^{\alpha/{(\alpha-1)}}
\eea
2)For harmonic oscillator[7]
\bea
E_{fho}=(\frac{\pi \hbar \gamma {D_{\alpha}}^{1/\alpha}q^{2/\gamma}}{2 B({1/\gamma},{{1/\alpha}+1})})^{\frac{\alpha \gamma}{\alpha+\gamma}}(n+\frac{1}{2})^
{\frac{\alpha \gamma}{\alpha+\gamma}}
\eea
\subsection{Fractal Properties }
The formula for fractal dimension of Continuous Time Random Walk(CTRW) is given by
$ 1+\beta(1-{1/\alpha})$[65]. For space fractional case it reduces to $2-{1/\alpha}$ ($\beta=1$).
To calculate the fractal dimension numerically we use a technique called detrended fluctuation analysis(DFA)[66,67,68] . It can be used to analyze time series that seem to be long-memory processes(diverging correlation time, for instance, a power law decaying autocorrelation function).
The DFA technique can be used to analyze signals which are non-stationary(changing over time). From the graph of the simulated paths, it is obvious that they are non-stationary and thus we use the DFA method to calculate the Hurst exponent.
The fractal dimension is calculated using the formula $D=2-H$ where H is the Hurst exponent[69]   
\newpage
\section{Results and discussions }
In section(3) we justify and prove the following formalism:
1) We extend the fractional heat equation to fractional space-time fractional FK formula.\\
2) For the Imprtance sampling we deduce  the expression for the  fractional version of the generator of the Fractional semigroup and we establish that the spectral problem corresponding to two generators are identical.\\
3) We deduce our version of the Generalized L$\acute{e}$vy path integral with appropriate L$\acute{e}$vy measure. \\
Fig 1 shows the trajectory for a space-time fractional case for a smaller step size.\\
Fig 2 shows a plot for a space-time fractional trajectory  for a bigger step size. \\  
Fig 3 is presented just to compare with the Gaussian trajectory. \\
Fig 4 is used to calculate the fractal dimension of the process generated with $\alpha=$ and $\beta=$. \\
Fig 5 shows that jumpsize increases with decrease in space index. \\
Fig 6 shows the corresponding wavefunctions for the harmonic oscillator for the fractional and integral space and time indices. \\
Fig 7 shows the signature of power law by having a fat tail for $\delta$ function potential.  \\
Fig 8 shows a  comparison of fat tail density in the fractional case with Gassian density for standard diffusion for $\delta$ function potential.\\ 
Fig 9 shows the depletion of density in the case of space-time fractional simple harmonic oscillator 
\subsection{Plots for various processes and densities}.\\ 
%Underneath we show the Gauusian trajectory:
\begin{figure}[h!]
\includegraphics[width=6in,angle=-0]{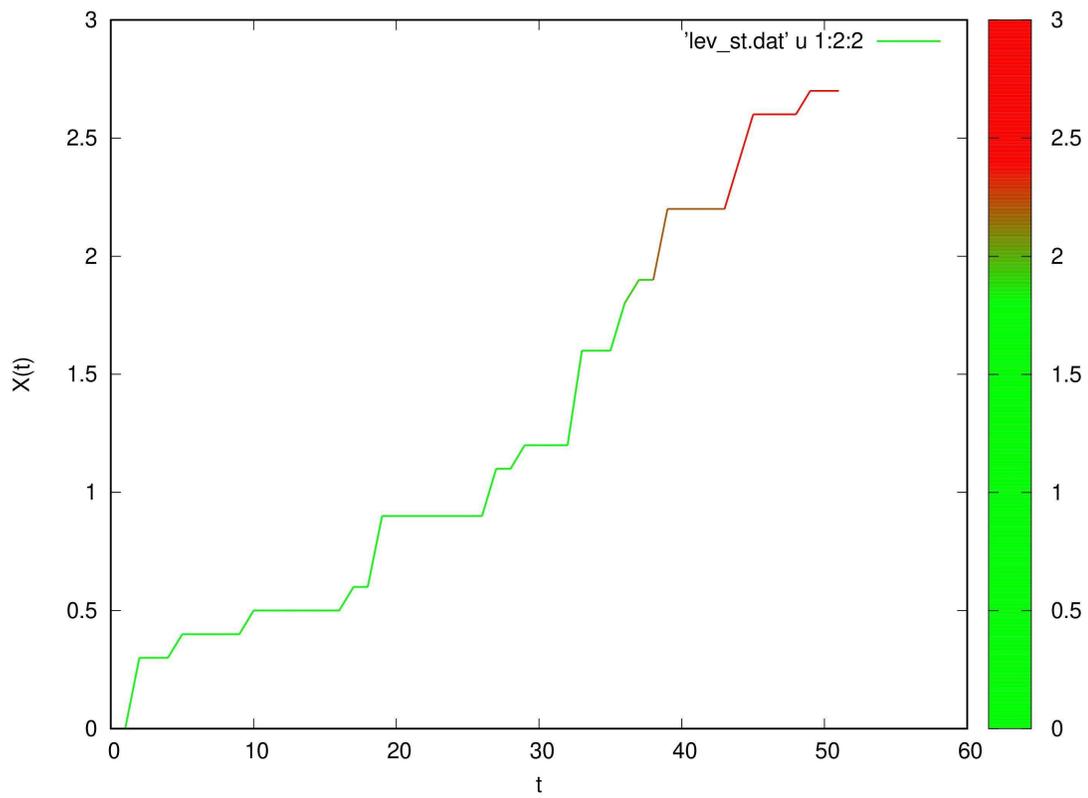}
\caption{Time series for the space-time fractiaonl process  for smaller number of steps}
\end{figure}
\begin{figure}[h!]
\includegraphics[width=4in,angle=-90]{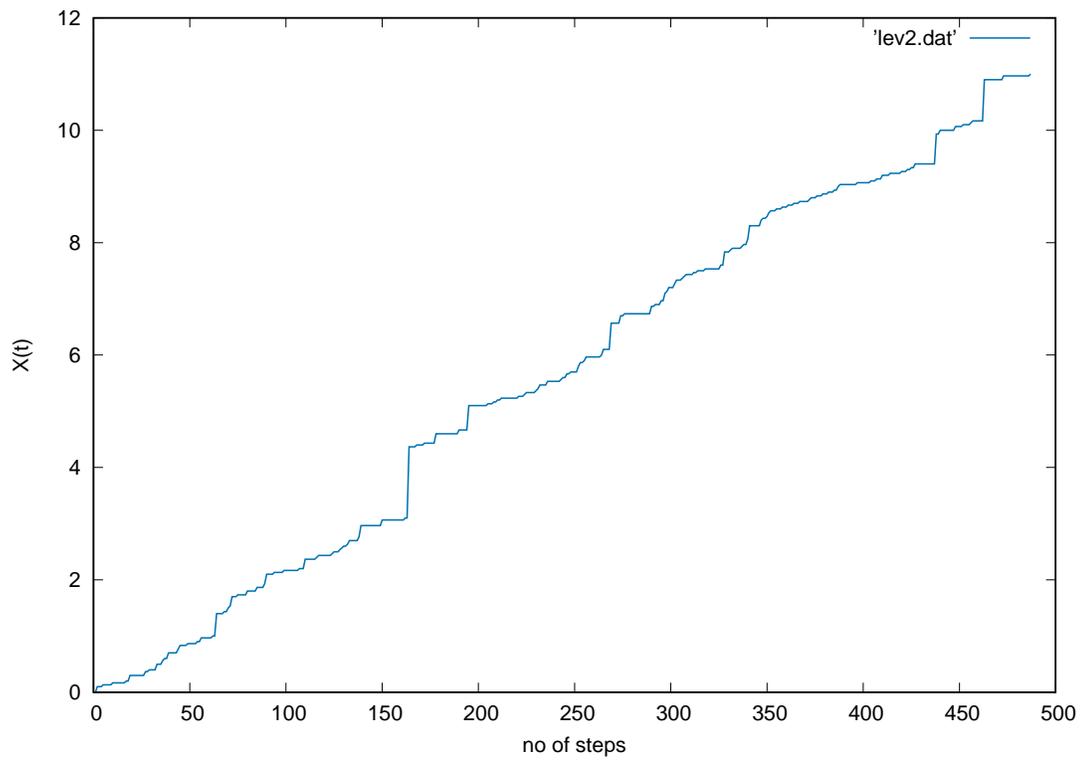}
\caption{ Time series for space-time fractional process for a larger number of steps}
\end{figure}
\clearpage
\newpage
\begin{figure}[h!]
\includegraphics[width=4in,angle=-90]{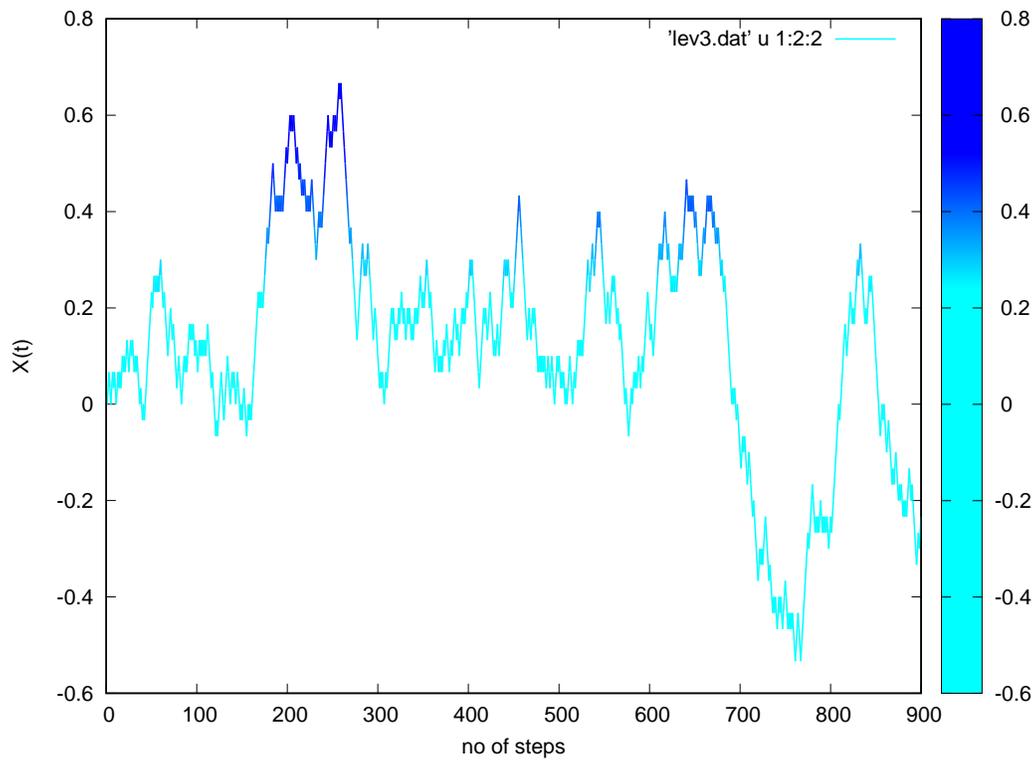}
\caption{Time series for a Gaussian trajectory }
\end{figure}
%\clearpage
\begin{figure}[h!]
\includegraphics[width=4in,angle=-90]{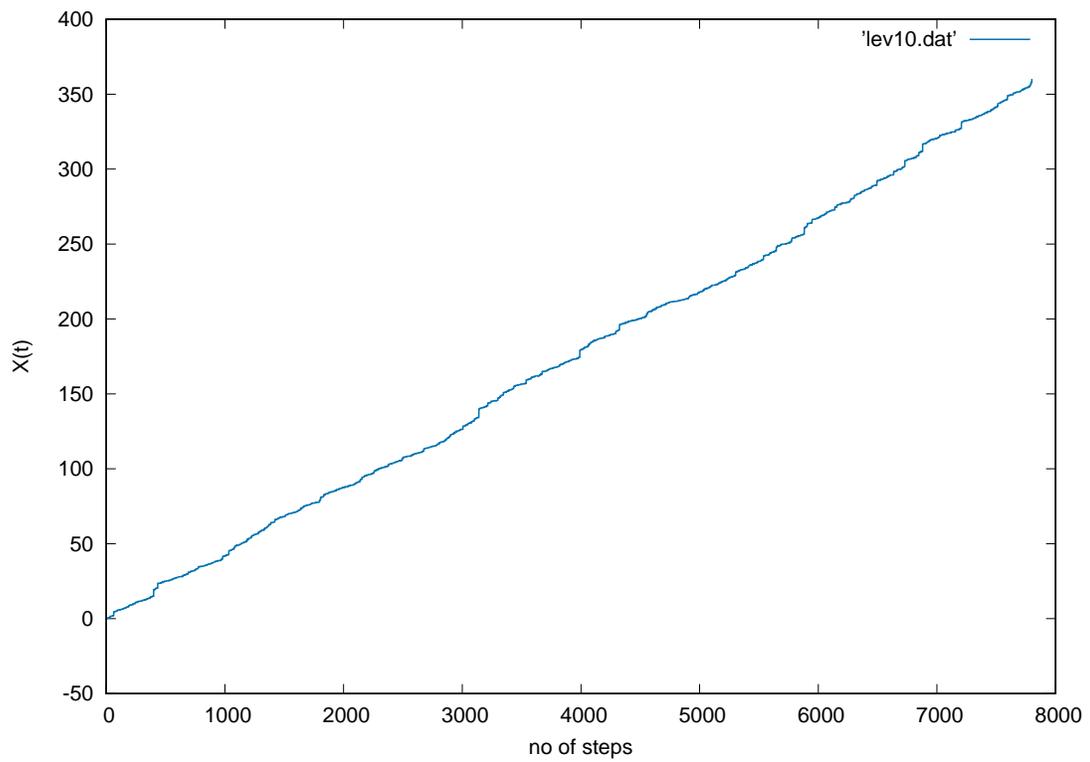}
\caption{A  space-time fractional trajectory for the  calculation of fractal dimension;$(\alpha=1.5,\beta=0.7)$}
\end{figure}
\begin{figure}[h!]
\includegraphics[width=4in,angle=-90]{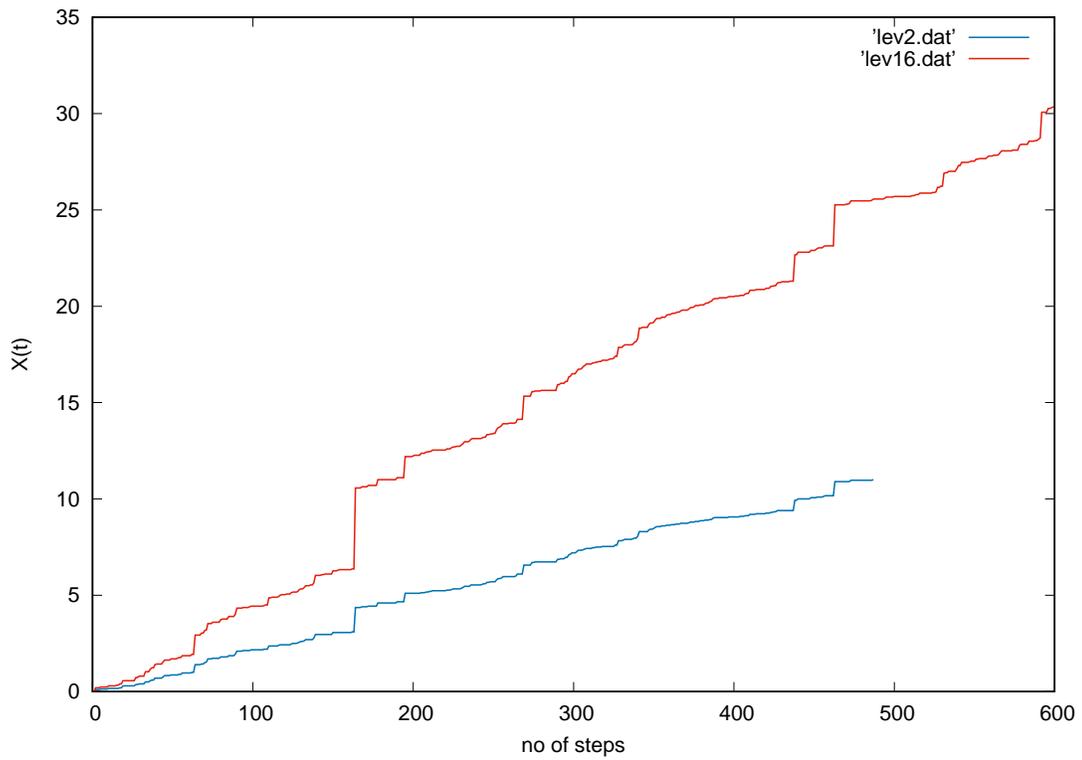}
\caption{ jumpsizes vs space index; the red graph with $\alpha=1.5$ has bigger jumps than blue graph with $\alpha=1.96$; resembles Fig-1 of Ref [39] }
\end{figure}
\begin{figure}[h!]
\includegraphics[width=4in,angle=-90]{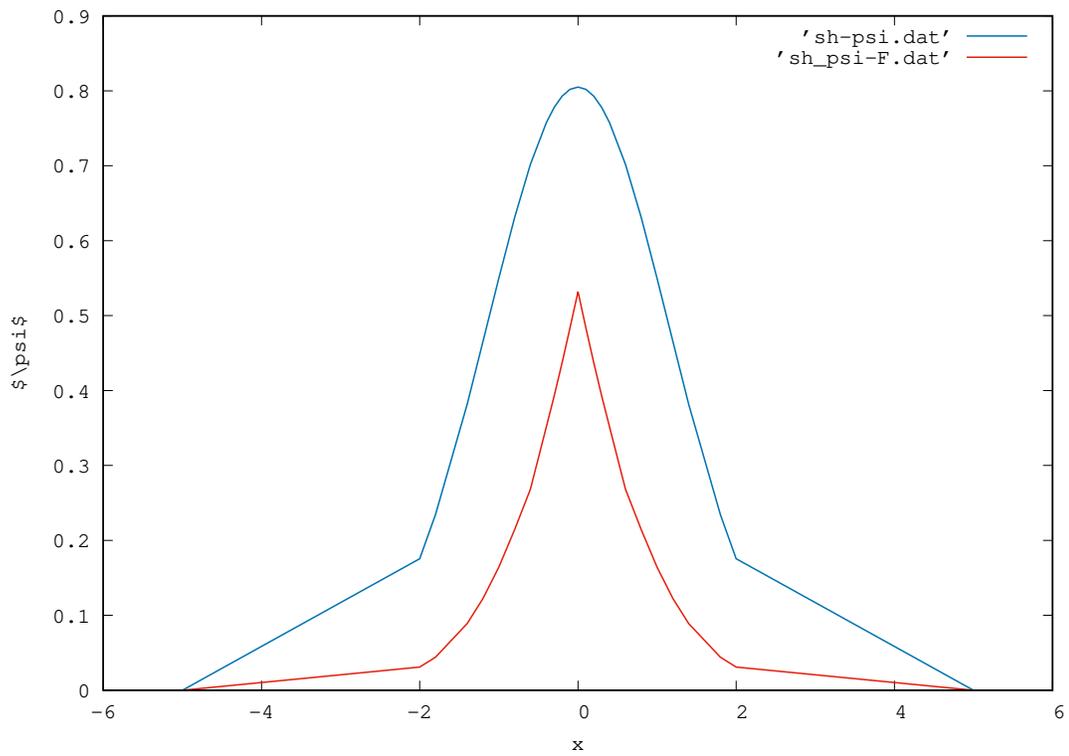}
\caption{A plot for the wave function for harmonic oscillator for fractional space and time indices; blue curve for $\alpha=2$ and red curve for $\alpha=1.5$}
\end{figure}
\begin{figure}[h!]
\includegraphics[width=6in,angle=-0]{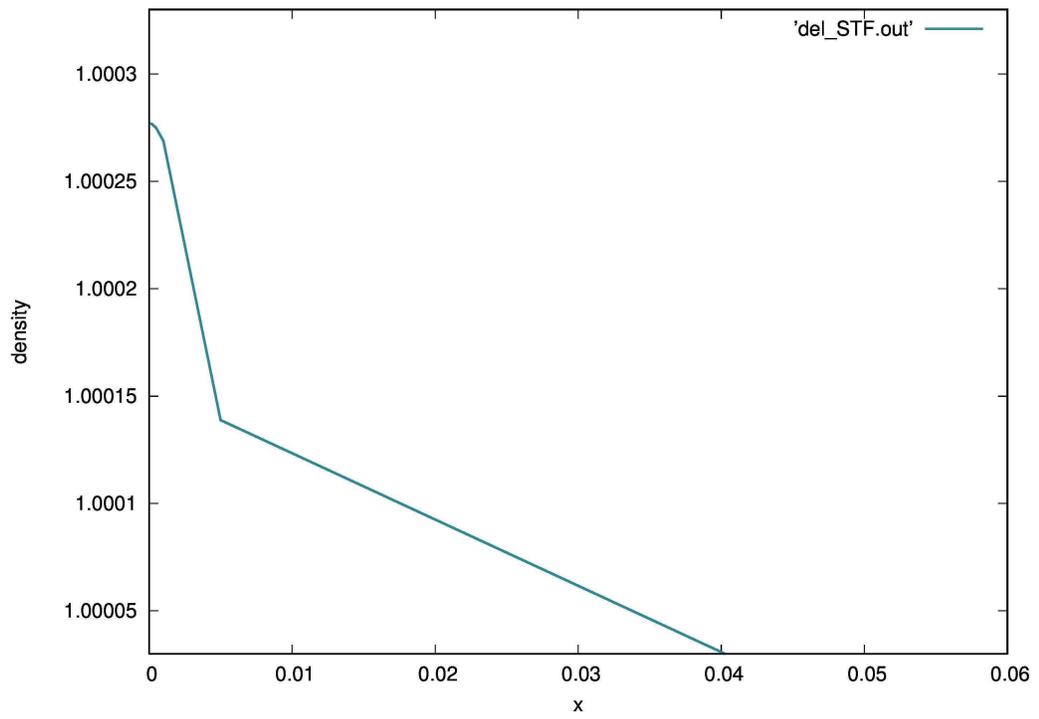}
\caption{A plot for the density for fractional space and time indices for $\delta $ function  potential}
\end{figure}
\begin{figure}[h!]
\includegraphics[width=4in,angle=-90]{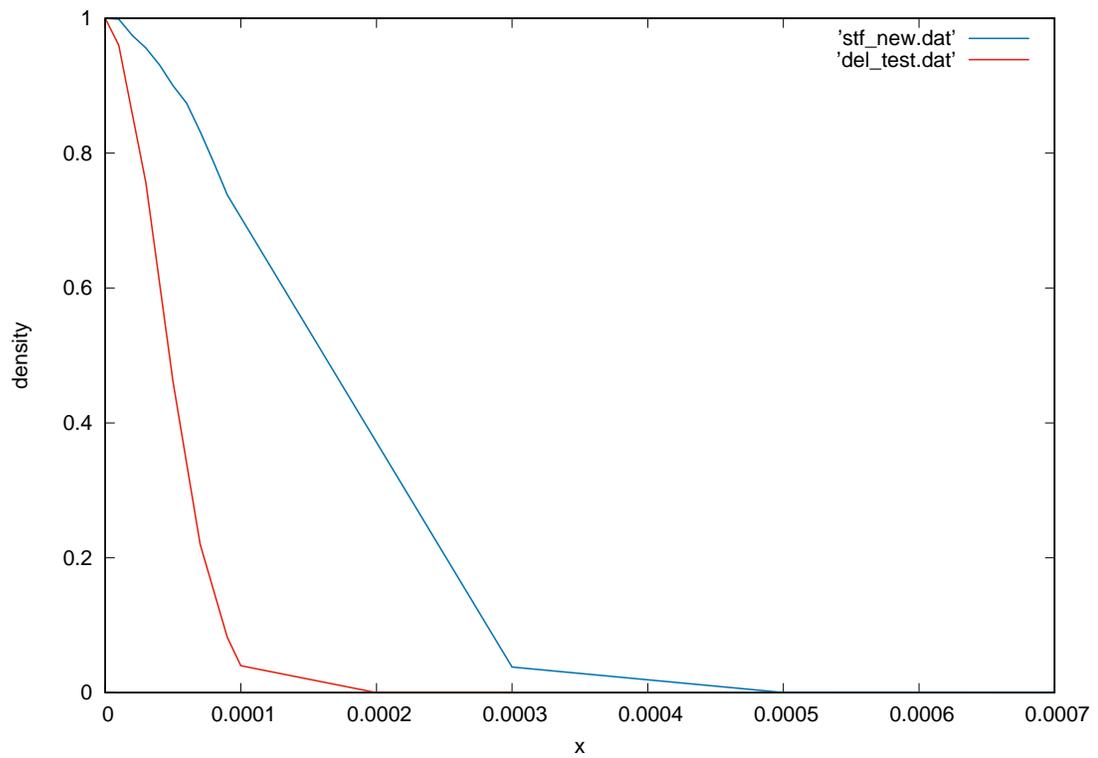}
\caption{density vs x; The graph with  $\alpha=1.96$,$\beta=0.98$(blue) shows Levy tail fatter than that of $\alpha=2$,$\beta=1 $ (red)}
\end{figure}
\begin{figure}[h!]
\includegraphics[width=6in,angle=-0]{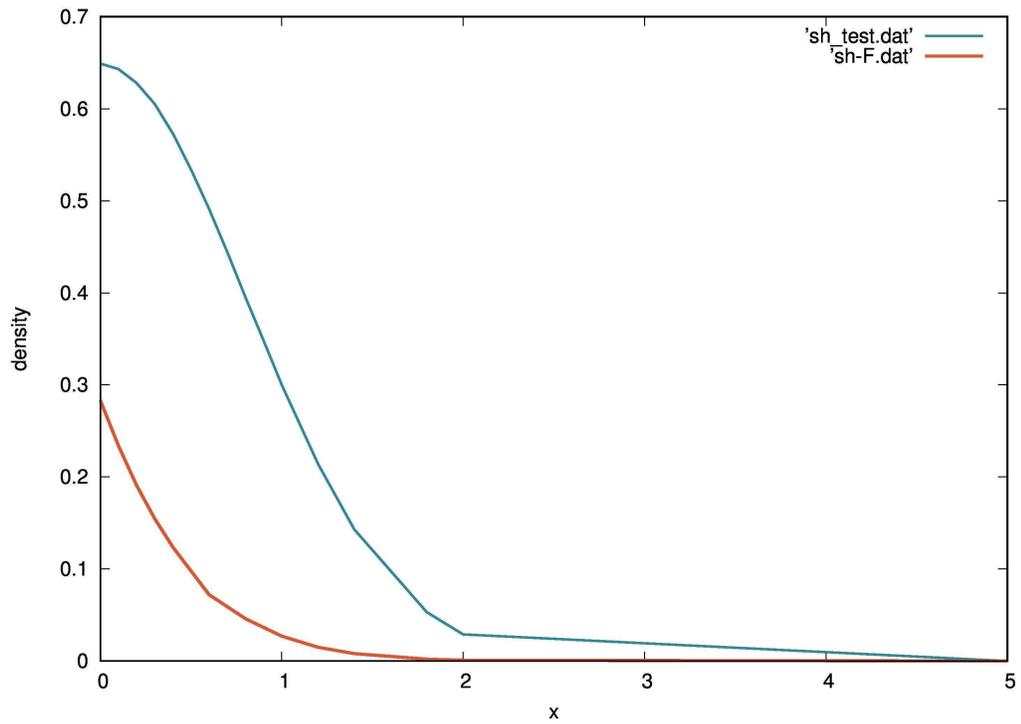}
\caption{A plot for the density of fractional(red) and integer harmonic oscillator(blue); shows similar depletion in the density as in Ref[29]in the case of fractional case.}  
\end{figure}
\clearpage
\begin{table}[h!]
\begin{center}
\caption{\bf Notation Table}
\begin{tabular}{ccc}
%\hline\hline
Notation/Phrase & Meaning \\
$\Delta$   &   Laplacian($ {\nabla}^2$)\\
$X(t)$     &   Brownian motion with a non-ergodic probabilistic measure or\\
           &    Wiener Measure\\
$Y(t)$     &   A stochastic process with an ergodic or stationary measure\\
$ \psi(x,t)$  &   The  trial function corresponding to mathematical ground state\\
$\phi_{0}$ &  The trial function for the quantum system\\
$\alpha $  &  The fractional  index of the space derivative \\
$\beta $   &  The fractional  index of the time derivative \\
$\kappa $  & $ \beta/\alpha $\\
$\delta $ & fractional index \\
$\delta(\vec{x})$ & $\delta$ function potential  \\  
%\hline\hline
\end{tabular}
\end{center}
\end{table}

\begin{table}[h!]
\begin{center}
\caption{\bf Numerical calculation of Fractal Dimensions(FD) of different processes} 
\begin{tabular}{llllll}
Data set & Hurst Exponent & FD(numerics) & FD(Theory) & $ \alpha $ & $ \beta $ \\
lev3 &  0.542259  & 1.457741 & 1.5   & 2 & 1       \\
lev10 & 0.7791264 & 1.220874 & 1.233 & 1.5 & 0.7 \\
\end{tabular}
\end{center}
\end{table}
\begin{table}[h!]
\begin{center}
\caption{\bf Comparison of Energy Eigenvalues by different methods }
\begin{tabular}{lllll}
 Systems &  Method & Space index &  Time index  & Energy eigenvalues \\
1(a) Fractinal simple & Fox H Finction &            &                    \\
Harmonic Oscillator &  representation & 1.5 & 1 & 0.62[Ref 6] \\
                                               
1(b) Fractional simple   &                 &     &   &      \\
 Harmonic Oscillator & FK simulation & 1.5 & 1 & 0.61(2)[This work] \\

1(c) Fractional simple  &Generalized                  &     &   &       \\
Harmonic Oscillator     &Feynman-Kac                  &     &   &      \\ 
                        &Simulation  & 1.5 & 1 & 0.690425[This work]  \\ 
2(a) $Schr\ddot{o}dinger$ &          &     &   &             \\
Equation             &                &     &  &              \\
with delta function  & Fox H function &      & &               \\
potential            & representation &  1.5 & 1 & -128.3000059[Ref 17] \\

2(b) $Schr\ddot{o}dinger$ &          &     &   &             \\
Equation             &                &     &  &              \\
with delta function  &                &      & &               \\
potential            & FK simulation &  1.5 & 1 & -128.359807(3) [This work]\\
\end{tabular}
\end{center}
\end{table}
\clearpage
In Table 1, we display the notational meaning for different quantities. 
In Table 2, we calculate the fractal dimention for the trajectories displayed in Fig 3 and 4.
In Table 3 we have presented our energy eigenvaues for space fractional harmonic osciallator and for  space fractional $Schr\ddot{o}dinger$ equation with $\delta function$ potential and compared the our results with other analytic calculations. For the case of space-time fractional case we do not have any numerical bench mark results. In this case we establish the correctness of our simulation by the follwing facts.\\
1) We calculate fractal dimensions which favorably compare wit the theoretical values.\\
2)The fat tail feature of the density shows that the CTRW was implemented successfully by generating random numbers appropriate for power-law charecterization of  L$\acute{e}$vy distribution.\\
3) The depletion in the density in the fractional case resembles the similar situation in Fig 1 of Ref[29] \\ 
In the appendix the scheme for simulating Brownian is shown. In the case of space-time fractional $Schr\ddot{o}dinger$ equation, the Brownian motion is replaced by L$\acute{e}$vy process  and is simulated by power law distriution.   
\clearpage
\section{Conclusions and Outlook}
Due to the complexity of nonlocal fractional space derivatives the fractional space-time diffusion equations with local potentials are difficult to handle analytically. On the contrary we found the numerical simulation based on path ntegration to be quite simple and easy to implement in such cases. We need to apply the Importance sampling technique for space fractional case with $\delta$ function  potential and other time independent potentials to test the power and worthwhileness of our method. We will continue to work on the problem and aspire to simulate the cases where the random jumps and the waiting times will not be independent to be able to solve  more realistic space-time fractional problems with heavy tails in the area of Physics, finance and biological sciences. We hope our method will inspire others to carry out similar investigations on other interesting fractional differential equations as an alternative to complicated fractional calculus.\\ 
\section{Acknowledgement}
One of the authors(SD) would like to thank  Alliance University for providing partial support for carrying the research work. SD would also like to thank Dr H. Kleinert of University of Berlin for suggesting the problem and Dr Andeizej Korzeniowski of The University of Texas
at Arlington, USA for many useful discussions. 
\clearpage
\newpage
\appendix
%\numberwithin{equation}{section}
\setcounter{equation}{0}
\renewcommand{\theequation}{A\arabic{equation}}
\section{Details of the Numerical Calculations}
The formalism described in section 2 can include any generalized potential [50] and valid for any arbitrary
dimension d (d=3N). To implement Eq(3) numerically, the 3N dimensional Brownian motion can be replaced by
properly scaled one dimensional random walks as follows [4,24,26]:
\ber
W(l)\equiv W(t,n,l)
& = & {w_1}^1(t,n,l),{w_2}^1(t,n,l),{w_3}^1(t,n,l)....\\ \nonumber
&   &                  .......{w_1}^N(t,n,l){w_2}^N(t,n,l){w_3}^N(t,n,l)
\eer
where
\bea
{w_j}^i(t,n,l)=\sum^l_{k=1}\frac{{\epsilon}^i_{jk}}{\sqrt n}
\eea
with ${w_j}^i(0,n,l)=0$
for $i=1,2,....,N$;$j=1,2,3$ and $l=1,2,.....,nt$. Here $\epsilon $ denotes the binomially distributed random variables which are
chosen independently and randomly with probability P for all i,j,k such that
$P({\epsilon}^i_{jk}=1)$=$P({\epsilon}^i_{jk}=-1)$=$\frac{1}{2}$. It is known
by an invariance principle [70] that for every $\nu$ and W(l)
defined in Eq.(A1) and Eq(A2)
\ber
\lim_{n\to\infty}P(\frac{1}{n}\sum^{nt}_{l=1}V(W(l)))\leq \nu \\ \nonumber
 =  P( \int\limits^t_0 V( X(s))ds)\leq\nu
 \,\,.
\eer
Consequently for large n,
\ber
P[ \exp(- \int\limits^t_0 V(X(s))ds)\leq\nu ] \\ \nonumber
 \approx  P [\exp(-\frac{1}{n}\sum^{nt}_{l=1}V(W(l)))\leq \nu]
\eer
Finally, by generating $N_{rep}$ independent realization $Z_1$,$Z_2$,....$Z_{N_{rep}}$ of
\bea
Z_m=\exp(-(-\frac{1}{n}\sum^{nt}_{l=1}V(W(l)))
\eea
and using the law of large numbers,with regard to Eq(A3), we conclude that
\bea
(Z_1+Z_2+...Z_{N_{rep}})/N_{rep}=Z(t)
\eea
is an approximation to Eq.(6)
Here $W^m(l), m=1,2,N_{rep}$ denotes the $m^{th}$ realization of W(l) out of
$N_{rep}$ independently run simulations. In the limit of large t and $N_{rep}$
this approximation approaches an equality, and forms the basis of a
computational scheme for the lowest energy of a many particle system with
a prescribed symmetry.

\end{document}